\documentclass[prd,11,twocolumn,floats,aps,epsfig,nofootinbib,amssymb]{revtex4-1}
\usepackage{graphicx}
\usepackage{epsfig}
\usepackage{subfigure}
\usepackage{dcolumn}
\usepackage{bm}
\usepackage{amssymb}
\usepackage{color}
\def\gsim{\lower0.5ex\hbox{$\:\buildrel >\over\sim\:$}}
\def\lsim{\lower0.5ex\hbox{$\:\buildrel <\over\sim\:$}}

\definecolor{Red}{rgb}{1,0,0}

\begin{document}


\title{Constraining Flavor Changing Interactions from LHC Run-2 \\ Dilepton Bounds with Vector Mediators}

\author{Farinaldo S. Queiroz$^1$}
\author{Clarissa Siqueira$^2$}
\author{Jos\'e W. F. Valle$^3$}
\email{queiroz@mpi-hd.mpg.de}

\affiliation{
$^1$ Max-Planck-Institut f\"ur Kernphysik, Saupfercheckweg 1, 69117 Heidelberg, Germany\\
$^2$ Departamento de F\'isica, Universidade Federal da Para\'iba, Caixa Postal 5008, 58051-970, Jo\~ao Pessoa - PB, Brazil\\
AHEP Group, Instituto de F\'isica Corpuscular – C.S.I.C./Universitat de Valencia
Edificio de Institutos de Paterna, C/Catedratico Jos\'e
Beltran, 2 E-46980 Paterna (Valencia) - SPAIN
}

\begin{abstract}
Within the context of vector mediators, is a new signal observed in flavor changing interactions, particularly in the neutral mesons systems $K^{0}-\bar{K}^{0}$, $D^{0}-\bar{D}^{0}$ and $B^0-\bar{B^0}$, consistent with dilepton resonance searches at the LHC? In the attempt to address this very simple question, we discuss the complementarity between flavor changing neutral current (FCNC) and dilepton resonance searches at the LHC run 2 at $13$~TeV with $3.2\, {\rm fb^{-1}}$ of integrated luminosity, in the context of vector mediators at tree level. Vector mediators, are often studied in the flavor changing framework, specially in the light of the recent LHCb anomaly observed at the rare B decay. However, the existence of stringent dilepton bound severely constrains flavor changing interactions, due to restrictive limits on the $Z^{\prime}$ mass. We discuss this interplay explicitly in the well motivated framework of a 3-3-1 scheme, where fermions and scalars are arranged in the fundamental representation of the weak SU(3) gauge group. Due to the paucity of relevant parameters, we conclude that dilepton data leave little room for a possible new physics signal stemming from these systems, unless a very peculiar texture parametrization is used in the diagonalization of the CKM matrix. In other words, if a signal is observed in such flavor changing interactions, it unlikely comes from a 3-3-1 model.
\end{abstract}

\maketitle

\section{Introduction}

The Standard Model (SM) has passed all precision tests thus far, and it is the best description of nature. Although, we need physics beyond the standard model so as to account for neutrino masses and dark matter. Many models that address these puzzles are plagued by flavor changing neutral current (FCNC) processes, which are, however, absent in the SM at tree-level, thanks to the GIM mechanism\footnote{The concept of minimal flavor violation has guided us at how to suppress new physics interactions \cite{Buras:2000dm,Buras:2003jf,D'Ambrosio:2002ex}.} \cite{Glashow:1970gm}. Therefore, precise measurement of flavor transition processes, such as those from neutral meson oscillations, $K^{0}-\bar{K}^{0}$, $D^{0}-\bar{D}^{0}$ and $B^{0}_d-\bar{B^0_d}$, which are forbidden in the SM at tree level, provide an excellent laboratory to test new physics models, due to lack of standard model background. Conversely, flavor changing charged currents, are overwhelmed by numerous W boson processes. 

That said, flavor changing neutral currents are often examined in the context of neutral vector gauge boson, $Z^{\prime}$. A multitude of Abelian and non-Abelian models predict the existence of extra neutral gauge bosons. Generally speaking they provide a straightforward 
cross-correlation among observables, such as FCNC and $Z^{\prime}$ at the LHC. Simplified models have become powerful tools in this endeavor, since they capture the main features of UV--complete models \cite{Allanach:2015gkd,Celis:2015eqs,Belanger:2016ywb,Altmannshofer:2016brv}. However, at the end of the day one needs a full theory to draw conclusive statements. In this attempt, we will address the complementarity between flavor changing neutral currents and dilepton resonance searches at the LHC, which refers to those with charged lepton pairs in the final state \cite{ATLAS1}, in the context of electroweak extensions of the SM, based on the $SU(3)_c \otimes SU(3)_L \otimes U(1)_N$ gauge group, shortly referred as 3-3-1 models.

3-3-1 models are self-consistent if there exists only three generations due to the combined effect of triangle gauge anomalies cancellations and QCD asymptotic freedom \cite{Singer:1980sw, Pisano:1991ee,Foot:1992rh,Montero:1992jk,Dias:2009au}. Moreover, the model furnishes a suitable environment for neutrino masses through see-saw mechanisms \cite{Mohapatra:1979ia,Minkowski:1977sc,Schechter:1980gr,Mohapatra:1980yp,
Lazarides:1980nt,  Schechter:1981cv,Keung:1983uu,Senjanovic:2010nq,Forero:2011pc,Dorame:2012zv,
Chen:2013fna,Boucenna:2014uma,Bonilla:2015eha,Valle:2016kyz}, dark matter \cite{deS.Pires:2007gi,Mizukoshi:2010ky,Alvares:2012qv,Queiroz:2013lca,
Kelso:2013nwa,Profumo:2013sca,Kelso:2014qka,Dong:2014esa,Dong:2014wsa,
Cogollo:2014jia,Martinez:2014ova,Martinez:2014ova,Martinez:2014rea,daSilva:2014qba,
Dong:2015rka,Martinez:2015wrp,Huong:2016ybt,Pires:2016vek}, explanation of the strong CP problem in the quark sector \cite{Dias:2002gg,Dias:2003zt},  first-order phase transitions \cite{Phong:2014yca,Phong:2014ofa,Long:2015qza}, lepton number violation processes \cite{Palcu:2007um,Dong:2008sw,Giang:2012vs,Hue:2013uw,Hue:2015fbb,Vien:2015wca,Machado:2016jzb,Fonseca:2016tbn}, and several others \cite{Schechter:1981bd,Cogollo:2007qx,Cogollo:2008zc,deS.Pires:2010fu,Alves:2011kc,Alves:2012yp,Caetano:2013nya,Hernandez:2013hea,Kelso:2013zfa,Cogollo:2013mga,Doff:2013wba,Machado:2013jca,Queiroz:2014zfa,Alves:2015pea,Buras:2015kwd,Doff:2015ukb,Doff:2015nru,Pires:2016dqq,Ferreira:2013nla,DeConto:2016osh,Hernandez:2016eod,Buras:2016dxz}. 3-3-1 models
are burden with FCNC interactions and they naturally arise at tree level in 331 model because one of the generations
has to transform differently from the other two, breaking the universality and leading to flavor changing interactions involving the new neutral gauge boson $Z^{\prime}$. In principle, there are also other sources of FCNC in the model involving the CP-even and -odd neutral scalar, but those are suppressed \cite{Cogollo:2012ek}.

In summary, in this work, we will investigate the degree of complementarity among flavor changing interactions and dilepton resonance searches at the LHC at 13 TeV with $3.2~fb^{-1}$ of integrated luminosity using A\-TLAS analysis \cite{ATLAS1}, which are linked to the $Z^{\prime}$ boson. Due to the paucity of relevant parameters dictating the results of both observables, and the fact that other 3-3-1 models feature mild changes in the $Z^{\prime}$ interactions with SM quarks, we are able to draw general conclusions which are applicable to many 3-3-1 models.

The paper is structured as follows: In Sec. II we briefly discuss the key aspects of the model relevant for our reasoning; In Sec. III, we obtain LHC bounds in the model using dilepton ATLAS 13 TeV data. In Sec. IV, we obtain FCNC stemming from the 3-3-1 model with right-handed neutrinos and outline the region which a FCNC signal can be seen in agreement with LHC data.

\section{THE MODEL}

The $SU(3)_c \otimes SU(3)_L \otimes U(1)_N$ gauge symmetry means that the fermions can be placed in the fundamental re\-presentation of $SU(3)_L$, i.e triplets. In order to reproduce the SM spectrum the SM doublet should be enclosed. The third component in the model is arbitrary and can vary from neutrinos, heavy neutrino fermions and even exotic charged leptons, depending on the quantum number assignments. There are two ways to incorporate right-handed neutrinos in the model. One can either add three
singlet right-handed neutrinos, or change the quantum numbers of the fermions in such way that right-handed neutrinos are embedded in the $SU(3)_L$ triplet. The latter scenario leads to an interesting and minimal model, which is the model we concentrate on, also shortly refereed as $331r.h.n$ firstly presented in \cite{Foot:1994ym,Hoang:1996gi,Hoang:1995vq}. Thus the lepton sector is,

\begin{equation}
f^a_L=\left(
\begin{array}{c}
\nu^a_l\\
e^a_l\\
(\nu^c_R)^a
\end{array}\right) \sim (1,3,-1/3), e_R^a \sim (1,1,-1), 
\end{equation} where $a=1,\, 2,\, 3$.

As for the hadronic sector, anomaly gauge cancellation demands that the first generation transforms as triplets under $SU(3)_L$, whereas the second and third one as anti-triplet as follows,
\begin{equation}\begin{array}{l}
Q_{1L} = \left(\begin{array}{c}
u_1\\
d_1\\
u_{1}^\prime\end{array}\right)_L \sim (3, 3, 1/3),\\
u_{1R} \sim (3, 1, 2/3),\
d_{1R} \sim (3, 1, -1/3),\ u^\prime_{1R} \sim (3, 1, 2/3),\\
Q_{iL} = \left(\begin{array}{c}
d_i\\
u_i\\
d_i^\prime\end{array}\right)_L \sim (3, \bar 3, 0), \\
u_{iR} \sim (3, 1, 2/3),\ d_{iR} \sim (3, 1, -1/3),
d_{iR}^\prime \sim (3, 1, -1/3),\end{array} \end{equation}
where $i = 2,3$, with $q^\prime$ being heavy exotic quarks with electric charges $Q(u^\prime_{1})=2/3$ and $Q(d^\prime_{2,3})=-1/3$.

One can straightforwardly check that all gauge anomalies
cancel with the above choice of gauge quantum numbers. In order to generate the fermion masses through the spontaneous symmetry
breaking mechanism three triplet scalars are needed. From a top-down approach, the scalar triplet $\chi$ with,

\begin{equation}
\langle \chi \rangle = \left(\begin{array}{c}
0\\
0\\
v_\chi\end{array}\right),
\end{equation} where $v_\chi$ is the vacuum expectation value of the neutral scalar responsible for breaking $SU(3)_L \otimes U(1)_N$ into $SU(2)_L \otimes U(1)_Y$, give rises to the exotic quark masses via the Yukawa Lagrangian,

\begin{equation}{\cal L}^{\chi}_{yuk}= \lambda_1 \bar Q_{1L} u^{\prime}_{1R} \chi
+ \lambda_{2ij} \bar Q_{iL} d^{\prime}_{jR} \chi^{*} + H.c., \end{equation}
where $\chi \sim (1, 3, -1/3)$. 

Then the $SU(2)\otimes U(1)_Y$ breaks into electromagnetism when two triplets $\rho, \eta$ acquire a vev with,

\begin{equation}
\langle \rho \rangle = \left(\begin{array}{c}
0\\
v_{\rho}\\
0\end{array}\right), \
\langle \eta \rangle = \left(\begin{array}{c}
v_{\eta}\\
0\\
0\end{array}\right),
\end{equation} giving rise to quark and charged lepton masses through the Yukawa lagrangian,

\begin{eqnarray}
{\cal L}_{Yuk} && = \lambda_{1a} \bar Q_{1L} d_{aR} \rho
+ \lambda_{2ia} \bar Q_{iL} u_{aR} \rho^{*} + G_{ab}
\bar f^a_L(f^b_L)^c \rho^{*} \nonumber\\
&&+ G^{'}_{ab} \bar f_L^a e_R^b \rho + \lambda_{3a} \bar Q_{1L} u_{aR} \eta +
\lambda_{4ia} \bar Q_{iL} d_{aR} \eta^{*} + H.c. \nonumber\\
\label{eq:yukawa2}
\end{eqnarray} with the scalar triplets transforming as $\rho \sim (1, 3, 2/3)$ and $\eta \sim (1, 3, -1/3)$. Moreover, the third term in Eq. (\ref{eq:yukawa2}) generates two degenerate masses to the neutrinos leaving one massless. This is problematic because one cannot explain the three mass differences observed in the neutrino oscillation data \cite{Kopp:2013vaa,Gonzalez-Garcia:2015qrr,Bergstrom:2015rba}. There are ways to generate neutrino masses in agreement with data through effect effective operators \cite{Queiroz:2010rj,Pires:2014xsa}, or by adding extra scalar to incorporate an inverse seesaw mechanism \cite{Dias:2012xp,Boucenna:2015zwa} with no prejudice to our reasoning which is concentrated on gauge interactions.

In this symmetry breaking pattern the $125$~GeV higgs mass is easily achieved and the SM gauge boson masses correctly obtained with,

\begin{eqnarray}
M_{W^\pm}^2    & = & \frac{1}{4}g^2v^2\,,\,M^{2}_{Z} = M_{W^\pm}^2/C^{2}_{W}, \nonumber \\
M^2_{Z^\prime} & = & \frac{g^{2}}{4(3-4S_W^2)}\left[4C^{2}_{W}v_{\chi}^2 +\frac{v^{2}}{ C^{2}_{W}}+\frac{v^{2}(1-2S^{2}_{W})^2}{C^{2}_{W}}\right ],\nonumber \\
M^2_{V^\pm}    & = & \frac{1}{4}g^2(v_{\chi}^2+v^2)\,,\,M^2_{U^0}= \frac{1}{4}g^2(v_{\chi}^2+v^2),
\label{massvec}
\end{eqnarray} where $Z^\prime$,$V^\pm$ and $U^0$, $U^{0 \dagger}$ are new gauge bosons predicted by the model, with $v^2= v_{\rho}^2+v_{\eta}^2$. We have now highlighted the key features of the model relevant to our reasoning, thus it is a good timing to discuss the collider phenomenology.

\begin{figure}[!h]
 \centering
 \includegraphics[width=0.5\columnwidth]{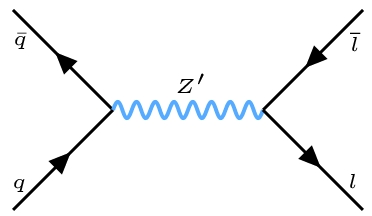}
 \caption{Feynman diagram relevant for dilepton production at the LHC.}
 \label{fig01}
\end{figure}

\section{Dilepton Resonance Searches at the LHC}

Heavy dilepton resonance searches at the LHC (see Fig.\ref{fig01}) have proven to be an effective channel to probe new physics models due to relatively good efficiencies/acceptance and well controlled background which comes mostly from Drell-Yann processes \cite{Jezo:2014wra,Jezo:2015rha,Klasen:2016qux}\footnote{ See \cite{delAguila:2010mx} for an excellent review about LEP-II limits}. Using 8 TeV center-of-energy and $20fb^{-1}$ of integrated luminosity ATLAS collaboration has placed restrictive limits on the mass of gauge bosons arising in some new physics models \cite{Aad:2014cka}, but an assessment particularly devoted to 3-3-1 models was performed in \cite{Salazar:2015gxa} ruling out $Z^{\prime}$ masses below $2.65$~TeV in the 3-3-1 model with right-handed neutrinos.

Here we take the dilepton results from LHC run II data at 13 TeV with $\mathcal{L}=3.2$ fb$^{-1}$ \cite{ATLAS1}, which has given rise to stringent limits on the $Z^{\prime}$ mass of several models including the sequential standard model reading $3.4$~TeV. For this type of analysis we have taken the background events using the results in \cite{ATLAS1}. The signal $pp \rightarrow Z^{\prime} \rightarrow l^+l^-$, where $l=e,\mu$, was simulated using MadGraph5 \cite{Alwall:2014hca,Alwall:2011uj} with the CTEQ6L parton distribution function \cite{Lai:2009ne} using efficiencies/acceptances described in \cite{Aad:2014cka}.

Similarly to previous analysis we selected the signal events using the cuts,

\begin{itemize}
\item  $E_T(e_1) > 30 \,{\rm GeV}, E_T(e_2) > {\rm 30 \,GeV}, |\eta_e| < 2.5$,
\item  $p_T(\mu_1) > 30 \,{\rm GeV}, p_T(\mu_2) > 30 \,{\rm GeV}, |\eta_{\mu}| < 2.5$,
\item $500 \,{\rm GeV} < M_{ll} < 6000 \,{\rm GeV}$,
\end{itemize}with $M_{ll}$ being the dilepton invariant mass.

These signals are peaked at the $Z^{\prime}$ mass, thus one can use cuts the dilepton invariant mass to discriminate signal from background. In summary, since no excess of events has been observed we can re-interpret ATLAS results to derive a limit on the $Z^{\prime}$ mass. Re-analyzing the ATLAS dilepton results we found $M_{Z^{\prime}} > 3$~TeV. It is important to stress that this limit is robust due to the paucity of relevant parameter in the analysis, namely the gauge couplings, which are fixed by the gauge symmetry of the model. With this limit in mind we now obtain the 3-3-1 contribution to FCNC processes in what follows.

\begin{figure}[!h]
 \centering
 \includegraphics[width=0.5\columnwidth]{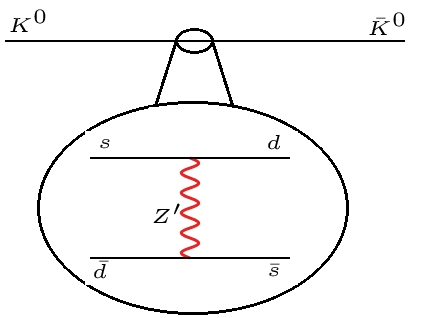}
 \caption{Diagram contributing to $K^0-\bar{K}^0$ mass difference in the 3-3-1 model with right-handed neutrinos.}
 \label{fig1}
\end{figure}

\begin{figure}[!h]
 \centering
 \includegraphics[width=0.5\columnwidth]{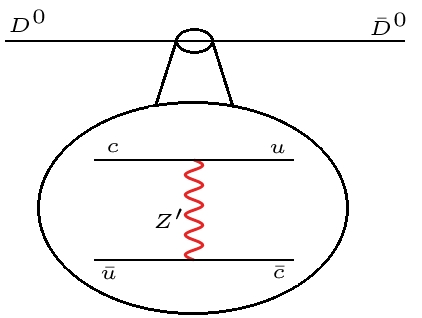}
 \caption{Diagram contributing to $D^0-\bar{D}^0$ mass difference in the 3-3-1 model with right-handed neutrinos.}
 \label{fig2}
\end{figure}

\begin{figure}[!h]
 \centering
 \includegraphics[width=0.5\columnwidth]{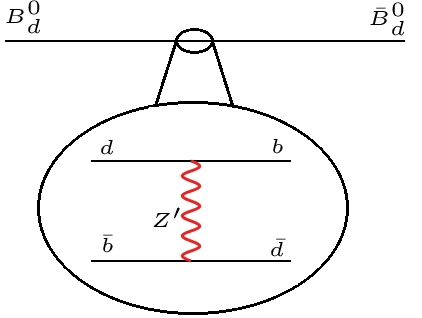}
 \caption{Diagram contributing to $B^0_d-\bar{B}^0_d$ mass difference in the 3-3-1 model with right-handed neutrinos.}
 \label{fig3}
\end{figure}

\section{FCNC in the 3-3-1}

All mesons are unstable, with the longest-lived lasting for only a few hundredths of a microsecond. Although no meson is stable, those of lower mass are nonetheless more stable than the most massive mesons, and are easier to observe in colliders. In particular the $K^0$ meson is a bound state composed of $d\bar{s}$, implying that kaons cannot be their own antiparticles. There must be then two different neutral kaons, differing by two units of strangeness, i.e. $K^0$ and $\bar{K}^0$ (see Fig. \ref{fig1}). The eigenstates which are obtained after mass diagonalization are known as Kaon long ($K_L$) and Kaon short ($K_S$) which yield opposite CP value, with $K_L$ decaying into three pions, and $K_S$ into two pions. Since $K_L$ is slightly heavier than three pion masses, its lifetime is much longer than the $K_S$. The physics of Kaon mixing is a explicit example of the importance of the CP symmetry in weak interactions.  The mass difference of these mesons is precisely measured to be $(\Delta m_K)=3.483 \times 10^{-12}$ MeV. In a similar vein, the mesons $D^0$ made of $c\bar{u}$ and $B_d^0$ composed of $d\bar{b}$ have mass difference $(\Delta m_D)=4.607 \times 10^{-11}$ MeV, $m_D=1865$~MeV and $(\Delta m_{B_d})=3.33 \times 10^{-10}$~MeV \cite{Garron:2011mz,Durr:2011ap,Beringer:1900zz} (see Figs.\ref{fig2}-\ref{fig3} and Table \ref{tableI}). Hence, new physics FCNC processes which might yield sizeable contributions to the mass differences above can be probed using these meson systems \footnote{See \cite{Misiak:1997ei,Barenboim:2000zz,Datta:2008qn} for relevant reviews.}. In the 3-3-1 model these FCNC processes that contribute to the mass difference of these meson systems surface through the neutral current mediated by $Z^{\prime}$ gauge boson (scalar contributions are dwindled). That said, in order to derive the 3-3-1 corrections to these mass differences in a pedagogic way, we need first to derive the neutral current in the 3-3-1 model. As in the SM the Z bosons does not mediated FCNC, only the $Z^{\prime}$ does through,

\begin{eqnarray}
\ensuremath{\mathcal{L}}^{Z^{\prime}}_{u} & = & \frac{g}{2C_{W}}\left(\frac{(3-4S_{W}^{2})}{3\sqrt{3-4S_{W}^{2}}}\right) \left[\bar{u}_{aL}\gamma_{\mu}u_{aL}\right] Z^{\prime}_{\mu} \nonumber \\
                                          &   & - \frac{g}{2C_{W}}\left(\frac{6(1-S_{W}^{2})}{3\sqrt{3-4S_{W}^{2}}}\right)\left[\bar{u}_{3L}\gamma_{\mu}u_{3L}\right] Z^{\prime}_{\mu},
\label{zprimau}
\end{eqnarray}

\begin{eqnarray}
\ensuremath{\mathcal{L}}^{Z^{\prime}}_{d} & = & \frac{g}{2C_{W}}\left(\frac{(3-4S_{W}^{2})}{3\sqrt{3-4S_{W}^{2}}}\right)
\left[\bar{d}_{aL}\gamma_{\mu}d_{aL}\right] Z_{\mu}^{\prime} \nonumber \\
                                          &   & - \frac{g}{2C_{W}}\left(\frac{6(1-S_{W}^{2})}{3\sqrt{3-4S_{W}^{2}}}\right)
\left[\bar{d}_{3L}\gamma_{\mu}d_{3L}\right] Z_{\mu}^{\prime},
\label{eq:FCNC1}
\end{eqnarray} with $a=1,\, 2,\, 3$, i.e. running through the three generations. 
Notice that Eqs. (\ref{zprimau}) and (\ref{eq:FCNC1}) are in the mass-eigenstate basis, but we need to move to the flavor basis in order to connect to meson observables using the transformations,

\begin{equation}
\left(\begin{array}{c}
u\\
c\\
t
\end{array}\right)_{L,R}
=
U_{L,R}
\left(\begin{array}{c}
u^{\prime}\\
c^{\prime}\\
t^{\prime}
\end{array}\right)_{L,R},
\left(\begin{array}{c}
d\\
s\\
b
\end{array}\right)_{L,R}
=
V_{L,R}
\left(\begin{array}{c}
d^{\prime}\\
s^{\prime}\\
b^{\prime}
\end{array}\right),
\label{misturaquarksMP}
\end{equation} where the matrices $U_{L,R}$ and $V_{L,R}$ are $3\times 3$ unitary and determine the Cabibbo-Kobayashi-Maskawa (CKM) matrix with $V_{CKM}=(U_{L})^{\dagger}(V_{L})$ ~\cite{cabibbo:1963yz,Kobayashi:1973fv,Agashe:2014kda}. Using this transformations one can find \cite{GomezDumm:1994tz,Long:1999ij,Benavides:2009cn},

\begin{eqnarray}
 \mathcal{L}^{K_0-\bar{K}_0}_{Z^\prime \, eff} &=& \frac{4 \sqrt{2}G_F C_W^4}{3-4 s_W^2}\frac{M_Z^2}{M_{Z^\prime}^2}|(V_L)_{31}^*(V_L)_{32}|^2|\bar{d}^{\prime}_{1L}\gamma_\mu d^{\prime}_{2L}|^2, \nonumber\\
 \mathcal{L}^{D_0-\bar{D}_0}_{Z^\prime \, eff} &=& \frac{4 \sqrt{2}G_F C_W^4}{3- 4s_W^2}\frac{M_Z^2}{M_{Z^\prime}^2}|(U_L)_{31}^*(U_L)_{32}|^2|\bar{u}^{\prime}_{1L}\gamma_\mu u^{\prime}_{2L}|^2, \nonumber\\
 \mathcal{L}^{B^0_d-\bar{B}^0_d}_{Z^\prime \, eff} &=& \frac{4 \sqrt{2}G_F C_W^4}{3-4 s_W^2}\frac{M_Z^2}{M_{Z^\prime}^2}|(V_L)_{31}^*(V_L)_{33}|^2|\bar{d}^{\prime}_{1L}\gamma_\mu d^{\prime}_{3L}|^2,\nonumber\\
 \label{eq:FCNC2}
\end{eqnarray} and consequently,

\begin{eqnarray}
 (\Delta m_K)_{Z^\prime} &=& \frac{4 \sqrt{2}G_F C_W^4}{3-4 S_W^2} \frac{M_Z^2}{M_{Z^\prime}^2} |(V_L)_{31}^*(V_L)_{32}|^2 f_K^2 B_K \eta_K m_K, \nonumber\\
 (\Delta m_D)_{Z^\prime} &=& \frac{4 \sqrt{2}G_F C_W^4}{3-4 S_W^2}\frac{M_Z^2}{M_{Z^\prime}^2}|(U_L)_{31}^*(U_L)_{32}|^2 f_D^2 B_D \eta_D m_D, \nonumber\\
 (\Delta m_{B_d})_{Z^\prime} &=& \frac{4 \sqrt{2}G_F C_W^4}{3-4 S_W^2}\frac{M_Z^2}{M_{Z^\prime}^2}|(V_L)_{31}^*(V_L)_{33}|^2 f_B^2 B_B \eta_B m_{B}, \nonumber\\ 
 \label{eq:FCNC3}
\end{eqnarray}with $G_F$ being the Fermi constant, $S_W (C_W)$ the sine (cossine) of the Weinberg angle, and $B_K,B_D,B_B$ the bag parameters, $f_K,f_D,f_B$ the decay constants, and $\eta_K,\eta_D,\eta_B$ the QCD leading order correction obtained in \cite{Misiak:1997ei,Barenboim:2000zz,Datta:2008qn}, and $m_K,m_D,m_B$ the masses of the mesons. In table \ref{tableI} we summarize the values of these parameters. 

We emphasize that the $Z^{\prime}$ does mediate FCNC in the 3-3-1 model because the hadronic generations do not transform identically under $SU(3)_L$. In Eqs. (\ref{zprimau})-(\ref{eq:FCNC3}) $u_a=u,\, d,\, t$ and $d_a=d,\, s,\, b$ for $a=1,\, 2,\, 3$ respectively, and $q^{\prime}$ representing the flavor eigenstate of a given quark.

\begin{table}[h]
{\color{blue} Input parameters}\\
\begin{tabular}{|c|}
\hline
$\Delta m_K=3.483 \times 10^{-12}$ MeV \\
 $m_K=497.614$ MeV  \\
 $\sqrt{B_K}f_K=135$ MeV \\
 $\eta_K=0.57$ \\
\hline
\end{tabular}

\begin{tabular}{|c|}
\hline
$\Delta m_D=4.607 \times 10^{-11}$~MeV  \\
 $m_D=1865$ MeV  \\
$\sqrt{B_D}f_D=187$ MeV  \\
$\eta_D=0.57$  \\
\hline
\end{tabular}

\begin{tabular}{|c|}
\hline
$\Delta m_{B_d} =3.33 \times 10^{-10}$ MeV \\
 $m_B=5279.5$ MeV \\
$\sqrt{B_B}f_B=208$ MeV  \\
$\eta_B=0.55$ \\
\hline
\end{tabular}
\caption{Limits on meson masses and numerical values for the bag parameters.}
\label{tableI}
\end{table}
\vspace{-0.5cm}
\begin{eqnarray}
&V&_{CKM} = 
\label{Vckm}\\
&& \left(
\begin{array}{ccc}
  0.97427 \scriptstyle{\pm 0.00014} & 0.22536 \scriptstyle{ \pm 0.00061} & 0.00355 \scriptstyle{\pm 0.00015} \\
 0.22522 \scriptstyle{\pm 0.00061} & 0.97343 \scriptstyle{\pm 0.00015} & 0.0414 \scriptstyle{\pm 0.0012} \\
  0.00886^{+0.00033}_{-0.00032} & 0.0405^{+0.0011}_{-0.0012} & 0.99914 \scriptstyle{\pm 0.00005}
 \end{array}\right). \nonumber
\end{eqnarray}

Now to compute the theoretical prediction from the 3-3-1 model to the mass difference systems under study as a function of the $Z^{\prime}$ mass, we simply need to plug into Eq.\ref{eq:FCNC3} the parameters summarized in Table \ref{tableI}, knowing the entries of  the quark mixing matrices $V^u_L$ and $V^d_L$. These entries are bound by the CKM matrix  (see Eq. \ref{Vckm}), which is reasonably well measured but the constraints on the individual entries of the matrices ($V^u_L$ and $V^d_L$) are loose \cite{Agashe:2014kda}. Therefore, one can work on two possible regimes which we name as {\it parametrization 1} and {\it parametrization 2}, which yield the strongest and weakest 3-3-1 contributions to FCNC processes respectively, while keeping the CKM matrix intact. In the {\it parametrization 1}, we found, 
$$
V_{L}=V_{R}=\left(
\begin{array}{ccc}
  0.97 & 0.23 &  0.0265598 \\
  0.23 & 0.97 &  0.096 \\
  0.043 & 0.089 &  0.995
 \end{array}\right)
$$ 
and,
$$
U_{L}=U_{R}=\left(
\begin{array}{ccc}
  0.89 & -0.45 & 0.00046 \\
  -0.45 & -0.89 &  0.06 \\
  0.0267 & 0.054 &  0.998 
 \end{array}\right),
$$ 
whereas for the {\it parametrization 2} we found,
$$
V_{L}=V_{R}=\left(
\begin{array}{ccc}
  0.965666 & -0.268135 &  0.0265598 \\
  -0.268135 & -0.968733 &  0.054013 \\
  0.0003757 & 0.0521882 &  0.99845 
 \end{array}\right)
$$ 
and,
$$
U_{L}=U_{R}=\left(
\begin{array}{ccc}
  0.877099 & -0.4759 &  0.00270598 \\
  -0.4739 & -0.8723 &  0.0106513 \\
  0.011237 & 0.020358 &  0.99999 
 \end{array}\right).
$$ 

\begin{figure}[!h]
 \centering
 \includegraphics[width=\columnwidth]{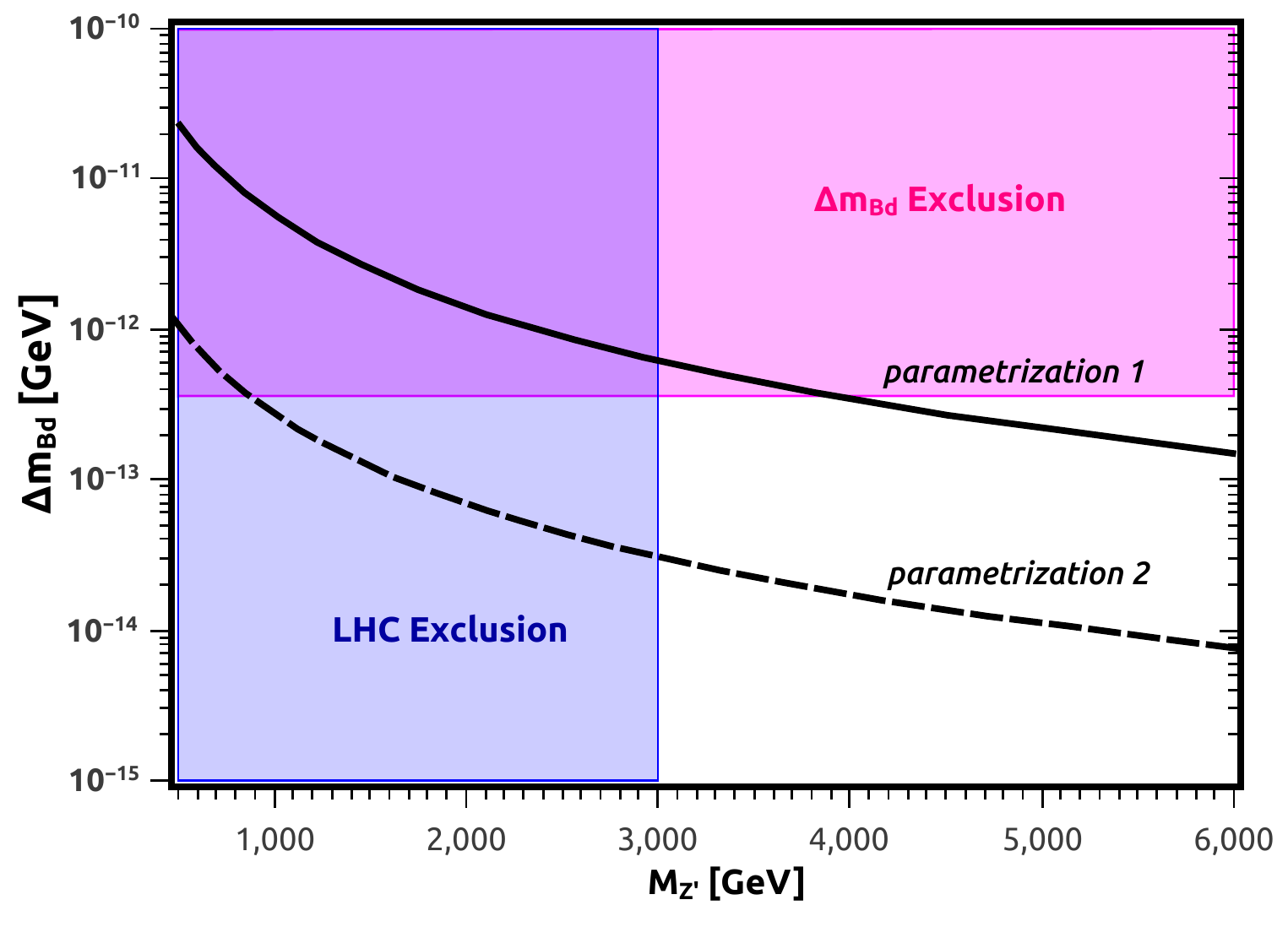}
 \caption{$\Delta m_{B_d}$ $\times$ $Z^\prime $ mass for two different parametrizations of the quark mixing matrices. The pink region is ruled out by constraints on $\Delta m_{B_d}$, wheres the shaded blue region indicate the exclusion limit on the $Z^\prime$ mass from LHC.}
 \label{figFCNC1}
\end{figure}

\begin{figure}[h]
 \centering
 \includegraphics[width=\columnwidth]{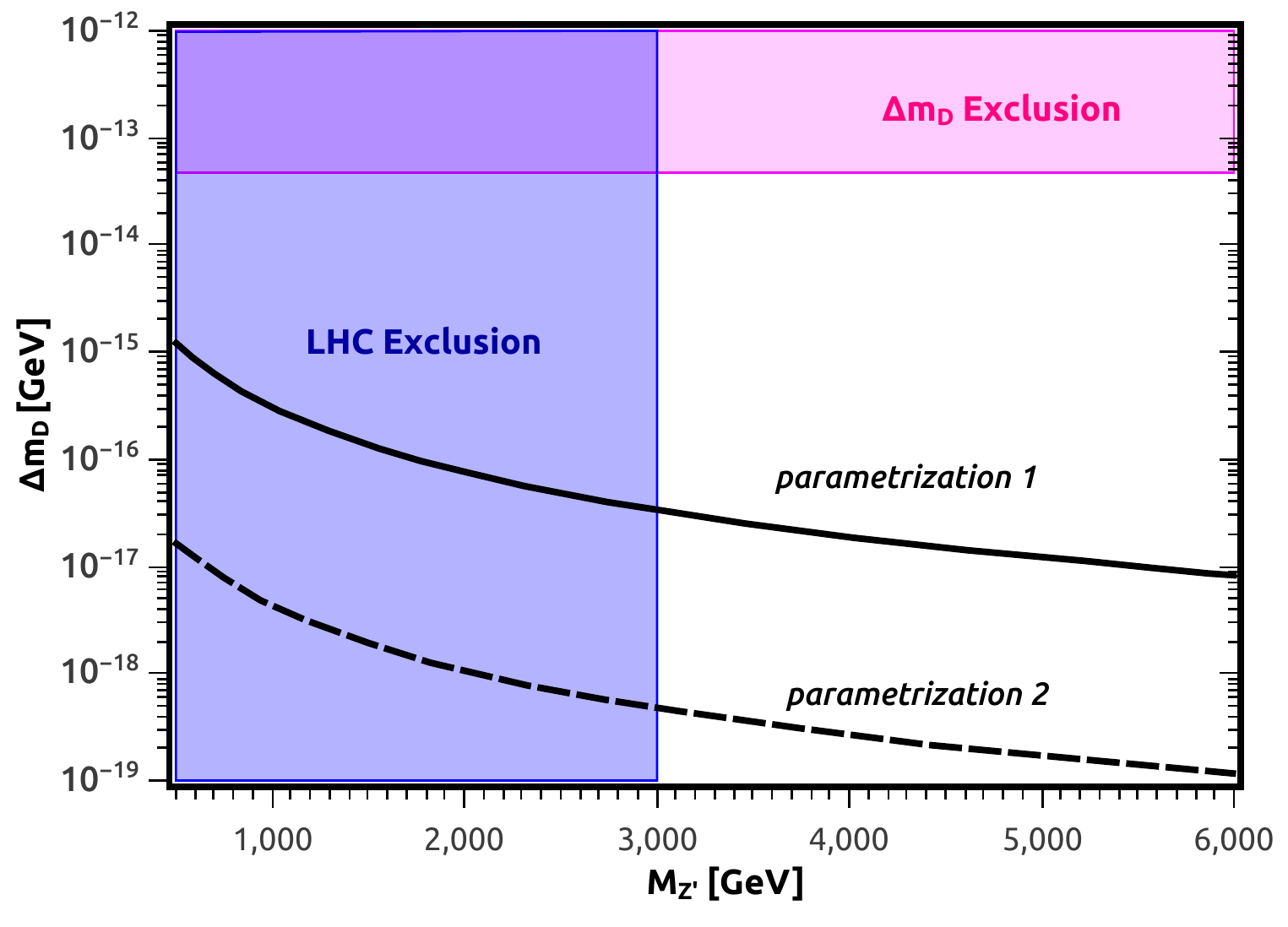}
 \caption{$\Delta m_D$ $\times$ $Z^\prime $ mass for two different parametrizations of the quark mixing matrices. The pink region is excluded by constraints on $\Delta m_D$ and the blue region is ruled out by the LHC limit on the $Z^\prime$ mass.}
 \label{figFCNC2}
\end{figure}

\begin{figure}[h]
 \centering
 \includegraphics[width=\columnwidth]{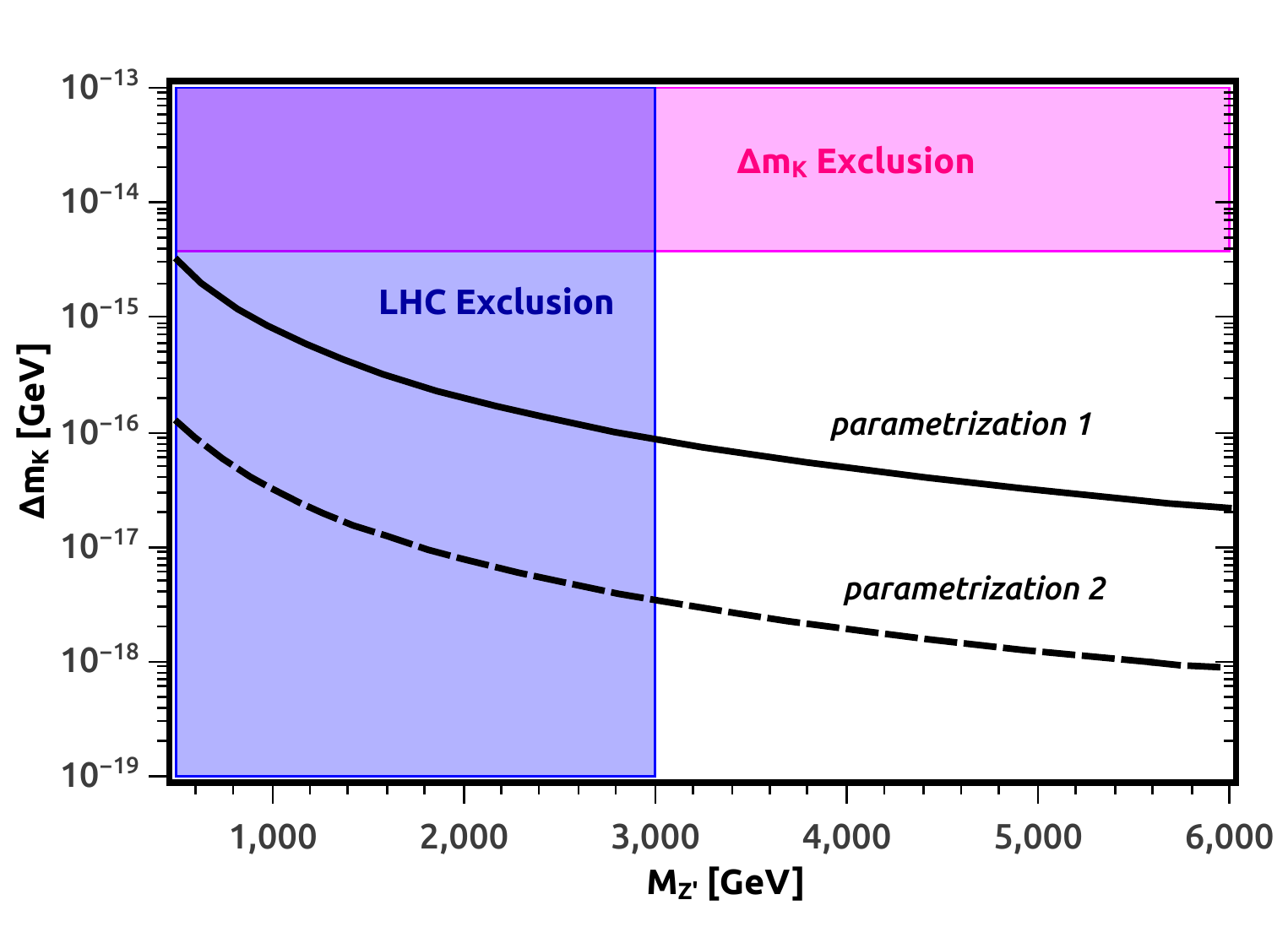}
 \caption{$\Delta m_K$ $\times$ $Z^\prime $ mass for two different parametrizations of the quark mixing matrices. The pink region is excluded by constraints on $\Delta m_K$ and the blue region is ruled out by the LHC limit on the $Z^\prime$ mass.}
 \label{figFCNC3}
\end{figure}

We have now collected all information needed to present the degree of complementarity between FCNC and dilepton searches at the LHC in the context of the vector mediator, $Z^{\prime}$ taking into account the uncertainties in which such constraints are subject to.

In Fig. \ref{figFCNC1} we show the 3-3-1 contribution to $\Delta m_{B_d}$  for {\it parametrizations 1-2} as a function of the $Z^{\prime}$ mass and we overlay in pink and blue the existing limits on the on the $B_d$ mass difference, and on the $Z^{\prime}$ mass coming from dilepton resonance searches at the LHC. Only using {\it parametrization 1} meson physics gives rise to a limit stronger than LHC one on the $Z^{\prime}$ mass. In other words, if in the near future a signal is observed in the $B_d$ system below the current limit, that would be consistent with LHC searches for a neutral vector boson. The 3-3-1 contribution to FCNC processes using {\it parametrization 2} is rather small, with LHC bound driving the limit on the $Z^{\prime}$ mass.

Moreover, in Figs.\ref{figFCNC2}-\ref{figFCNC3} we see that the 3-3-1 corrections to the mass difference of the $K^0$ and $D^0$ mesons is quite dwindled. Thus LHC rules out any possibility for a possible signal in the foreseeable future coming from the 3-3-1 model, since the LHC limits on the $Z^{\prime}$ mass is very stringent and robust, which reads $M_{Z^{\prime}} > 3$~TeV. In other words,  dilepton data from the LHC leaves basically no window for a possible FCNC signal in these systems to come from a 3-3-1 model unless a parametrization which enhances the 3-3-1 corrections to FCNC processes is advocated as it occurs in the {\it parametrization 1} for the $B_d$ meson system.




\section{Conclusion}

We have investigated the degree of complementarity between FCNC in the neutral mesons systems $K^{0}-\bar{K}^{0}$, $D^{0}-\bar{D}^{0}$ and $B^0_d-\bar{B^0_d}$ in the context of vector mediators, using the 3-3-1 model with right-handed neutrinos as framework. Our goal was to assess the possibility of explaining a possible FCNC signal in these systems having in mind the stringent limits stemming from dilepton resonance searches at the LHC. After briefly presenting the model we derived the $13$~TeV LHC $3.2 fb^{-1}$ limit on the $Z^{\prime}$ mass which reads 3 TeV. Then we proceeded to the 3-3-1 corrections to the mass differences of the three mesons above. We found that the 3-3-1 contributes appreciably only the $B^0_d$ mass difference. Using two different parametrizations, one that enhances, {\it parametrization 1} and other that suppresses {\it parametrization 2} the 3-3-1 contribution to the latter, we concluded that bounds on the $Z^{\prime}$ rising from dilepton resonance searches generally impose much stronger limits than FCNC ones. Conversely, a small window for a signal in the $B_d$ system exists if {\it parametrization 1} is used. Therefore, if a FCNC signal is seen in these mesons systems in the foreseeable future, unless a parametrization very similar to {\it parametrization 1} is advocated, the 3-3-1 model cannot not offer a feasible solution.

\begin{acknowledgments}
The authors thank the anonymous referee for several suggestions. This work is supported by the Spanish grants FPA2014-58183-P, Multidark CSD2009-00064, SEV-2014-0398 (MINECO) and PROMETEOII/2014/084 (GVA). CS acknowledges support from CNPq, Brazil. JV thanks Carlos Pires for hospitality at the Departamento de F\'isica, Universidade Federal da Para\'iba in Jo\~ao Pessoa, Brazil. FSQ thanks UNICAMP and UFABC for the hospitality during final stages of this project. FSQ is specially grateful to Alex Dias and Vanidia Dias for discussions and hospitality. 
\end{acknowledgments}

\bibliography{darkmatter}

\end{document}